\documentclass[11pt,a4paper]{article}
\usepackage{amsfonts}
\usepackage{graphicx}
\usepackage{amsmath}
\usepackage{hyperref}
\usepackage{enumerate}
\usepackage{amsmath,amssymb}
\usepackage{slashed,mathrsfs}
\usepackage{feynmp,subfigure}
\usepackage{verbatim,graphicx}
\usepackage[sub,ovp]{psfragx}
\usepackage{overpic}

\RequirePackage{mathrsfs} \RequirePackage[sc]{mathpazo}
\RequirePackage{wasysym} \RequirePackage{setspace}
\newcommand{\be}{\begin{equation}}
\newcommand{\ee}{\end{equation}}
\newcommand{\bea}{\begin{eqnarray}}
\newcommand{\eea}{\end{eqnarray}}
\input{tcilatex}
\begin{document}

\date{}
\title{ \rightline{\mbox{\small
{}}}\textbf{\ On Exchange-Correlation Energy in DFT Scenarios}}
\author{A. Belhaj$^{1}$\thanks{%
a-belhaj@um5r.ac.ma} and S. E. Ennadifi$^{2}$\thanks{%
ennadifis@gmail.com } \thanks{%
Authors in alphabetical order} \thanks{%
Corresponding author: ennadifis@gmail.com} \\
{\small $^{1}$ESMaR, Faculty of Sciences, Mohammed V University in Rabat,
Rabat, Morocco}\\
{\small $^{2}$LHEP-MS, Faculty of Sciences, Mohammed V University In Rabat,
Rabat, Morocco }\\
}
\maketitle

\begin{abstract}
Motivated by the considerable importance of material properties in modern
condensed matter physics research, and using techniques of the $N_{e}$%
-electron systems in terms of the electron density $n_{\sigma e}\left(
r\right) $ needed to obtain the ground state energy $E_{e0}$ in Density
Functional Theory scenarios, we approach the Exchange-Correlation energy $%
E_{xc}\left[ n_{\sigma e}(r)\right] $ by considering the interelectronic
position corrections $\Delta r_{x}^{\uparrow \uparrow ,\uparrow \downarrow
}=\lambda _{x}\left\vert \delta r^{\uparrow \uparrow }-\delta r^{\uparrow
\downarrow }\right\vert $ and $\Delta r_{c}^{e_{i}e_{j\neq i}}=\lambda
_{c}\left\vert r-r^{\prime }\right\vert ^{-\left( N_{e}-1\right) ^{-1}}$
corresponding to the spin and the Coulomb correlation effects, respectively,
through the electron-electron potential energy. Exploiting such corrections,
we get approximate expressions for the exchange $E_{x}\left[ n_{\sigma e}%
\right] $ and the correlation $E_{c}\left[ n_{\sigma e}\right] $ functional
energies which could be interpreted in terms of magnetic and electric dipole
potential energies associated with the charge density $n_{\sigma e}\left(
r\right) $ described by inverse-square potential behaviors. Based on these
arguments, we expect that such an   obtained Exchange-Correlation functional
energy could be considered in the Local Density Approximation functional as
an extension to frame such interelectronic effects.


\ \ \ \ \ \ \ \ \ \ \ \emph{\ \ \ \ \ \ \ \ \ \ }.
\end{abstract}

\newpage

\section{Introduction}

Today, it is well-known that Quantum Mechanics (QM) is the successful theory
for describing the behaviour of physical objects at the microscopic scales 
\cite{1,2,3,4}. However, for general potential energies, approximations are
needed. For the electronic structure calculations, conventional methods try
to obtain approximate solutions to the Schr\"{o}dinger equation of
many-interacting electrons $N_{e}$ moving in an external electrostatic
potential generated by atomic nuclei. These methods confront severe
restrictions related to the structure of the ensued many-electron
wavefunction $\Psi _{\sigma e}\left( x_{i}\right) $ and the quickly
increasing calculation stress, which makes the characterization of huge
systems turn into a very coasting task. The most notorious self-coherent
field approach in the matter of the electronic structure computation is the
Hartree-Fock theory\ (HF) \cite{5,6,7}. Later, despite the fact that Density
Functional Theory (DFT) has its origins in the Thomas--Fermi model \cite{8,9}%
, it was first put on a firm theoretical footing in the Kohn-Sham (KS)
equations \cite{10,11}. As initially demonstrated by Hohenberg and Kohn (HK)
in \cite{11}, the total energy and hence all other properties of the system
can be characterized. Nevertheless, the rigorous ground state energy can be
yielded by DFT only if the Exchange-Correlation (XC) energy is known, and
since it is always hard to determine the associated physical quantity
expressions, models are developed and are employed in an approximate way for
real systems. Among others, the most used approximations in DFT are the
Local Density Approximation (LDA) and Generalized Gradient Approximation
(GGA) \cite{12,13}. In spite of the expanded computational techniques, the
advancement of precise and flexible XC energy in DFT scenarios remains a
defy \cite{14,15,16}. Another possible route for dealing with such an
unknown energy is via the consideration of singular potentials which have
been largely discussed in QM \cite{17,18,19,20}.

In this work, we reconsider the study of the XC energy in DFT scenarios. In
the second section, we give a brief review of the DFT computational scheme
using the electron density $n_{\sigma e}$ $\left( r\right) $ for getting the
ground state energies $E_{e0}$ of many-electron systems $N_{e}$. In the
third section, we approach the XC functional energy $E_{xc}\left[ n_{\sigma
e}\right] $ by considering the interelectronic position corrections $\Delta
r_{x}^{\uparrow \uparrow ,\uparrow \downarrow }=\lambda _{x}\left\vert
\delta r^{\uparrow \uparrow }-\delta r^{\uparrow \downarrow }\right\vert $
and $\Delta r_{c}^{e_{i}e_{j\neq i}}=\lambda _{c}\left\vert r-r^{\prime
}\right\vert ^{-\left( N_{e}-1\right) ^{-1}}$ corresponding to the spin and
the Coulomb correlation effects, respectively, via the electron-electron
potential energy. Then,  we discuss the obtained approximate expressions for
the exchange $E_{x}\left[ n_{\sigma e}\right] $ and the correlation $E_{c}%
\left[ n_{\sigma e}\right] $ energies and interpret them in terms of known
electric and magnetic dipole potential energies, which could be considered
as an extension of the LDA functional to comprise such interelectronic
effects. The last section is devoted to conclusion and open questions.

\section{Many-electron system and electron density}

To start \footnote{%
Atomic units where $\hbar =m_{e}=\frac{e^{2}}{4\pi \varepsilon _{0}}=1$ are
used.}, we consider a system of $N_{e}$ electrons with the space-spin
coordinates $x_{i}=\left( r_{i},\sigma _{i}\right) $. Such mixed coordinates
generate the structure $\mathbb{R}^{3}\times \mathbb{Z}_{2}\left( \uparrow
\downarrow \right) $. In the non-relativistic and the Born--Oppenheimer
approximations \cite{12}, the ground state electronic energy can be
expressed as

\begin{equation}
E_{e0}=\min_{\Psi _{\sigma e}\left( x_{i}\right) }\left\langle \Psi _{\sigma
e}\left( x_{i}\right) \right\vert \widehat{H}_{e}\left\vert \Psi _{\sigma
e}\left( x_{i}\right) \right\rangle \text{, \ }\Psi _{\sigma e}\left(
x_{i}\right) =\Psi _{\sigma e0}\left( x_{i}\right)  \label{eq1}
\end{equation}%
where $\widehat{H}_{e}$ is the corresponding electronic Hamiltonian and $%
\Psi _{\sigma e0}\left( x_{i}\right) $ is the many-electron wavefunction
that minimizes the energy expectation value the energy expectation value $%
E_{e}=\left\langle \widehat{H}_{e}\right\rangle =\left\langle \Psi _{\sigma
e}\left( x_{i}\right) \right\vert \widehat{H}_{e}\left\vert \Psi _{\sigma
e}\left( x_{i}\right) \right\rangle $ under the normalization and the
anti-symmetrization conditions of the wavefunction. However, this is still a
hard task as it belongs to the spacetime computational complexity $C_{\Psi
_{\sigma e}\left( x_{i}\right) }$ of the issue which ascents exponentially
with the electron number $C_{\Psi _{\sigma e}\left( x_{i}\right) }\sim
e^{N_{e}}$ \cite{14} \footnote{%
Known as the Van Vleck/Orthogonality Catastrophe.}. As an overcoming of this
computational problem, instead of the many-electron wavefunction $\Psi
_{\sigma e}\left( x_{i}\right) $, the one-electron density $n_{\sigma
e}\left( r\right) $ with $n_{\sigma e}\left( r\rightarrow +\infty \right)
\rightarrow 0$ is used as a fundamental variable such as

\begin{equation}
n_{\sigma e}\left( r\right) =N_{e}\sum\limits_{\sigma =\sigma _{2}}^{\sigma
_{N_{e}}}\int_{\mathbb{R}^{3N_{e}}}\left\vert \Psi _{\sigma e}\left( \left(
r,\sigma _{1}\right) ,x_{2},\ldots ,x_{N_{e}}\right) \right\vert
^{2}dr_{2}\ldots dr_{N_{e}}  \label{eq2}
\end{equation}%
with certain properties related to the total electron number $N_{e}$
conservation and to electron charge density. In effect, this will reduce the
dimensionality $D_{\Psi _{\sigma e}\left( x_{i}\right) }$ of the studied
system by $\times N_{e}^{-1}$ as $D_{\Psi _{\sigma e}\left( x_{i}\right)
}=3N_{e}\overset{\times N^{-1}}{\rightarrow }D_{n_{\sigma e}\left( r\right)
}=3$. With this dimensionality reduction,  the computationnality of the theory
scales only as $C_{n_{\sigma e}\left( r\right) }\sim N_{e}^{3}$ and thus the
solution of the problem becomes computationally achievable, even for large
systems. Having this electron density variable at hand, the deal now is that
wether or not it is possible to rewrite the electronic energy $E_{e}$ as a
functional of the the electron density. In fact, this deal seems to be
achieved partially through the external potential energy term from the
electron-nuclei interaction. In particular, we have 
\begin{equation}
E_{e}=E_{e}\left[ n_{\sigma e}\right] =T_{e}\left[ n_{\sigma e}\right]
+V_{ee}\left[ n_{\sigma e}\right] +V_{en}\left[ n_{\sigma e}\right] =F_{e}%
\left[ n_{\sigma e}\right] +\sum\limits_{\sigma =\sigma _{1}}^{\sigma
_{N_{e}}}\int_{\mathbb{R}^{3}}v_{ext}\left( r\right) n_{\sigma e}\left(
r\right) dr  \label{eq3}
\end{equation}%
where the first term refers to the electron kinetic energy, the second one
is the interelectronic repulsion energy, and the last one represents the
external energy or the electron-nucleus attraction energy with $%
v_{ext}\left( r\right) =-\sum\limits_{\alpha =1}^{N_{\alpha }}\frac{%
Z_{\alpha }}{\left\vert r_{i}-R_{\alpha }\right\vert }$, where $Z_{\alpha }$
and $R_{\alpha }$ are the charges and the positions of the $N_{n}$ nuclei,
is the corresponding potential. The term $F_{e}\left[ n_{e}\right] $ is the
universal functional \footnote{%
Independent of the external potential.}. According to the early
approximation attempts conducted by TF \cite{8,9}, the total electronic
energy could be expressed as 
\begin{equation}
E_{e}^{TF}\left[ n_{\sigma e}\right] =\frac{3}{10}\left( 3\pi ^{2}\right)
^{2/3}\sum\limits_{\sigma =\sigma _{1}}^{\sigma _{N_{e}}}\int_{\mathbb{R}%
^{3}}n_{\sigma e}^{5/3}\left( r\right) dr+\frac{1}{2}\sum\limits_{\sigma
=\sigma _{1}}^{\sigma _{N_{e}}}\int_{\mathbb{R}^{3}\times \mathbb{R}^{3}}%
\frac{n_{\sigma e}\left( r\right) n_{\sigma e}\left( r^{\prime }\right) }{%
\left\vert r-r^{\prime }\right\vert }dr\text{\ }dr^{\prime
}+\sum\limits_{\sigma =\sigma _{1}}^{\sigma _{N_{e}}}\int_{\mathbb{R}%
^{3}}v_{ext}\left( r\right) n_{\sigma e}\left( r\right) dr  \label{eq4}
\end{equation}%
where the second term is the interelectronic classical Coulomb interaction
energy $V_{ee}^{classic}\left[ n_{\sigma e}\right] $. This remains only an
approximation approach seen that it does not include the XC energy of
electrons, though the extended version of TF theory by Dirac known as TFD
tried somehow to take into account the exchange energy. This is a central
subject of DFT which will be reconsidered in what follows.

The starting point of DFT is the HK theorems \cite{11} stating, roughly,
that due to the functional relationship between the ground state electron
density of a system and an arbitrary external potential $v_{ext}\left(
r\right) \rightarrow n_{\sigma e}\left( r\right) $ and $n_{\sigma e}\left(
r\right) \rightarrow v_{ext}\left( r\right) +cst$, the total energy of a
system is a unique functional of the electron density $E_{e}^{HK}=E_{e}^{HK}%
\left[ n_{\sigma e}\left( r\right) \right] $ and the lowest energy is given
if and only if the input density is the true ground state density $%
E_{e0}^{HK}=E_{e}\left[ n_{\sigma e0}\left( r\right) \right] $. Thus, the
total energy and all other properties of the system can be inferred.
According to this, the exact ground state energy for the total energy (\ref%
{eq3}) can be then expressed as 
\begin{equation}
E_{e0}^{HK}=E_{e}\left[ n_{\sigma e0}\left( r\right) \right]
=\min_{n_{e}\left( r\right) }\left\{ F_{e}\left[ n_{\sigma e}\right]
-\sum\limits_{\sigma =\sigma _{1}}^{\sigma _{N_{e}}}\int_{\mathbb{R}%
^{3}}v_{ext}\left( r\right) n_{\sigma e}\left( r\right) dr\right\} ,\text{ \ 
}n_{\sigma e}\left( r\right) =n_{\sigma e0}\left( r\right) .  \label{eq5}
\end{equation}%
It follows that the knowledge of the unknown universal functional $F_{e}%
\left[ n_{e}\right] $ would be sufficient to determine the total energy of
the system as well as its properties in the ground state, hence the use of
approximations leading to the transition towards the consideration of a
system of independent electrons evolving in external potential. This allows
to re-express the total energy functional in the KS form \cite{10,11} as 
\begin{equation}
E_{e}^{KS}=E_{e}\left[ n_{\sigma e}\right] =T_{s}\left[ n_{\sigma e}\right]
+V_{KS}\left[ n_{\sigma e}\right]  \label{eq6}
\end{equation}%
where $T_{s}\left[ n_{\sigma e}\right] $ is the non-interacting kinetic
energy. The KS potential energy $V_{KS}$ reads as 
\begin{eqnarray}
V_{KS}\left[ n_{\sigma e}\right] &=&V_{en}\left[ n_{\sigma e}\right] +\text{%
\ }V_{H}\left[ n_{\sigma e}\right] +E_{xc}\left[ n_{\sigma e}\right]  \notag
\\
&=&\sum\limits_{\sigma =\sigma _{1}}^{\sigma _{N_{e}}}\int_{\mathbb{R}%
^{3}}v_{ext}\left( r\right) n_{\sigma e}\left( r\right) dr+\frac{1}{2}%
\sum\limits_{\sigma =\sigma _{1}}^{\sigma _{N_{e}}}\int_{\mathbb{R}%
^{3}\times \mathbb{R}^{3}}\frac{n_{\sigma e}\left( r\right) n_{\sigma
e}\left( r^{\prime }\right) }{\left\vert r-r^{\prime }\right\vert }dr\text{\ 
}dr^{\prime }+E_{xc}\left[ n_{\sigma e}\right] \text{\ }  \label{eq7}
\end{eqnarray}%
where $V_{H}\left[ n_{\sigma e}\right] $ is the Hartree energy, and $E_{xc}%
\left[ n_{\sigma e}\right] $ represents the XC energy which is 
\begin{equation}
E_{xc}\left[ n_{\sigma e}\right] =\left( T_{e}\left[ n_{\sigma e}\right]
-T_{s}\left[ n_{\sigma e}\right] \right) +\left( V_{ee}\left[ n_{\sigma e}%
\right] -V_{H}\left[ n_{\sigma e}\right] \right) .  \label{eq8}
\end{equation}%
In this way, the first contribution $\left( T_{e}\left[ n_{\sigma e}\right]
-T_{s}\left[ n_{\sigma e}\right] \right) $ describes the corrections to the
kinetic energy while the second one $\left( V_{ee}\left[ n_{\sigma e}\right]
-V_{H}\left[ n_{\sigma e}\right] \right) $ indicates all non-classical
corrections to the electron-electron repulsion \cite{5,6}. With the KS
potential energy \ref{eq7}), the corresponding KS effective monoelectronic
potential in which electrons move is given by 
\begin{eqnarray}  \label{VKS}
\upsilon _{KS}\left( r\right) &=&\delta _{n_{\sigma e}\left( r\right)
}V_{KS} \left[ n_{\sigma e}\right] =\upsilon _{ext}\left( r\right) +\upsilon
_{H}\left( r\right) +\upsilon _{xc}\left( r\right)  \label{eq9} \\
&=&\upsilon _{ext}\left( r\right) +\sum\limits_{\sigma =\sigma _{1}}^{\sigma
_{N_{e}}}\int_{\mathbb{R}^{3}}\frac{n_{\sigma e}\left( r^{\prime }\right) }{%
\left\vert r-r^{\prime }\right\vert }dr^{\prime }\text{\ }+\upsilon
_{xc}\left( r\right)  \notag
\end{eqnarray}%
where $\upsilon _{H}\left( r\right) $ is the Hartee potential and $\upsilon
_{xc}\left( r\right) $ is the XC potential. This provides the primary
theoretical background needed to build an effective single-particle design
which enables the ground state density calculation and the energy of systems
of interacting electrons. Roughly, using the KS orbital equations \cite%
{10,11}, the the real interacting energy of the system (\ref{eq3}) can be
expressed in the following form 
\begin{equation}
E_{e}=\sum\limits_{i=1}^{N_{e}}\varepsilon _{i}-V_{H}\left[ n_{\sigma e}%
\right] \text{\ }-\sum\limits_{\sigma =\sigma _{1}}^{\sigma _{N_{e}}}\int_{%
\mathbb{R}^{3}}\upsilon _{xc}\left( r\right) n_{\sigma e}\left( r\right)
dr+E_{xc}\left[ n_{\sigma e}\right] .  \label{eq10}
\end{equation}%
To provide an explicit expression, this requires the determination of the XC
energy. In what follows, a special emphasis will be put on such an energy.

\section{Exchange-Correlation energy approach}

\subsection{Approximation approach}

Ignoring consideration of the electron-electron interactions radically
reduces the precision of computational prediction and results. Thus, the
additional pure quantum mechanical XC energy, given in (\ref{eq14}), that
can describe the pure interelectronic interaction must be implemented.
Finding the most accurate and efficient XC energy is one of the challenges
in DFT. The search for expressions of XC functionals is the primary running
defy within DFT with many years of inability and hard achievements. Since
defining expressions for quantities of a real system is always a difficult
task, the relevant expressions have been developed and exploited in an
approximate manner for real systems. In DFT, amongst the most-widely used
ones are the LDA, GGA approximations as well as derived methods which are
based on a non-local approach.e.g., meta GGA and HM-GGA \cite{13,14,15,16}.
These methods, depending on the approaches used (homogeneous electron gas
for LDA and spatial varying electron density for GGA), allow for obtaining
approximate results with certain precision degrees. Unfortunately, the exact
density functionals of the XC effects are not yet known and thus they are
still not included in the traditional DFT methods.

Here, we follow a different path to approach the XC energy in a heuristic
way based on an interelectronic position correction\textbf{\ }which could be
considered as an extension of the LDA functional to include all possible
spatial interelectronic effects.  It is denoted that such an energy
has been introduced in (\ref{eq8}) as the sum of the error made in using a
non-interacting kinetic energy and the error made in treating the
electron-electron interaction classically. As it is known, it is usually
useful to handle the complexities of the physical systems in parts rather
than try to solve them all in one go. Actually, in the spirit of LDA, we
begin by splitting this energy into an exchange part and into a correlation
part as 
\begin{equation}
E_{xc}\left[ n_{\sigma e}\right] =E_{x}\left[ n_{\sigma e}\right] +E_{c}%
\left[ n_{\sigma e}\right]  \label{eq11}
\end{equation}%
where the first exchange energy part $E_{x}$ is used to deal with  a relation
between electrons having parallel/antiparallel spin. However, the second
correlation energy part $E_{c}$ is used to deal with relation between
electrons due to the Coulomb effect. In essence, these two parts are both,
mostly, associated with, but separate concepts from, the interelectronic
potential energy $V_{ee}\left[ n_{\sigma e}\right] $. In this approach, we
could restrict our deal with such an energy to the potential energy part of
its definition in (\ref{eq8}) as

\begin{equation}
E_{xc}\left[ n_{\sigma e}\right] \simeq V_{ee}\left[ n_{\sigma e}\right]
-V_{ee}^{classic}\left[ n_{\sigma e}\right] \equiv V_{ee}\left[ n_{\sigma e}%
\right] -V_{H}\left[ n_{\sigma e}\right] .  \label{eq12}
\end{equation}%
This could constitute an inspiration in the attempt to approximate the
expressions of these energies. Concretely, this leads to the fact that one
has to deal essentially with the electronic spatial position parameter $r$.
Indeed, being a direct consequence of the Pauli exclusion principle, the
exchange energy part is a spin correction 
\begin{equation}
\Delta r_{x}^{\uparrow \uparrow ,\uparrow \downarrow }=f_{x}\left( \delta
r^{\uparrow \uparrow },\delta r^{\uparrow \downarrow }\right)  \label{eq13}
\end{equation}%
according to which electrons with same spin prefer to stay apart, whereas
ones with opposite spin tend to get closer. For the correlation energy part,
it accounts for the Coulomb interaction correction

\begin{equation}
\Delta r_{c}^{e_{i}e_{j\neq i}}=f_{c}\left( r_{i},r_{j}\right)  \label{eq14}
\end{equation}%
originating from the correlation between the spatial positions of electrons,
rather than the average field of electrons $n_{\sigma e}\left( r\right) $,
due to the interelectronic potential energy. In other words, the spatial
position of an electron $r_{i}$ is not random and is correlated to the
spatial positions $r_{j\neq i}$ of the surrounding $\left( N_{e}-1\right) $
electrons, but not the average field of electrons. The displacement of one
electron is always correlated to the surrounding electrons. Such an energy
has an effect on both the interelectronic potential energy $V_{ee}\left[
n_{\sigma e}\right] $ as well as the kinetic energy of the system $T_{s}%
\left[ n_{\sigma e}\right] $. With this at hand, the XC energy could be
split as follows 
\begin{equation}
E_{xc}\left[ n_{\sigma e}\right] =E_{x}^{\uparrow \uparrow ,\uparrow
\downarrow }\left[ n_{\sigma e}\right] +E_{c}^{e_{i}e_{j}}\left[ n_{\sigma e}%
\right] .  \label{eq15}
\end{equation}%
For the position correction $\Delta r_{x}^{\uparrow \uparrow ,\uparrow
\downarrow }$ due to the spin interaction of electrons of
parallel/antiparallel spins, and the position correction $\Delta
r_{c}^{e_{i}e_{j}\neq i}$ due to the electron position correlation to the
positions of the $N_{e}-1$ surrounding electrons, we could propose, inspired
from the inverse Coulomb potential behaviors and dimensionality reduction
along with some physics limit requirements, the possible expressions for the
exchange $f_{x}$ (\ref{eq13}) and the correlation $f_{c}$ (\ref{eq14})
functions in terms of the electron positions as follows 
\begin{eqnarray}
f_{x}\left( \delta r^{\uparrow \uparrow },\delta r^{\uparrow \downarrow
}\right) &=&\lambda _{x}\left\vert \delta r^{\uparrow \uparrow }-\delta
r^{\uparrow \downarrow }\right\vert \equiv \lambda _{x}\left\vert \delta
r^{\uparrow \uparrow -\uparrow \downarrow }\right\vert  \label{eq16} \\
f_{c}\left( r_{i},r_{j}\right) &=&\frac{\lambda _{c}}{\left\vert
r_{i}-r_{j}\right\vert ^{\frac{1}{N_{e}-1}}}  \label{eq17}
\end{eqnarray}%
where $\lambda _{x}$ and $\lambda _{c}$ are some specific dimension-full
factors related to the exchange and the correlation energies, respectively,
which will be discussed later on. The exponent in (\ref{eq17}) could be
identified with the dimension of the ordinary projective space $\mathbb{CP}^{%
{N_{e}}-1}$ supported by the geometric interpretation of $n_{\sigma e}$. In
this interelectronic interaction reasoning the whole XC energy according to (%
\ref{eq12}) can be thought of, roughly, as the difference between the
spatially corrected interelectronic repulsion energy and the uncorrected one
as 
\begin{equation}
E_{xc}\left[ n_{\sigma e}\right] =\frac{1}{2}\sum\limits_{\sigma =\sigma
_{1}}^{\sigma _{N_{e}}}\int_{\mathbb{R}^{3}\times \mathbb{R}^{3}}\left( 
\frac{n_{\sigma e}\left( r\right) n_{\sigma e}\left( r^{\prime }\right) }{%
\left\vert r-r^{\prime }\right\vert +\left( \lambda _{x}\left\vert \delta
r^{\uparrow \uparrow -\uparrow \downarrow }\right\vert +\lambda
_{c}\left\vert r-r^{\prime }\right\vert ^{-\left( N_{e}-1\right)
^{-1}}\right) }-\frac{n_{\sigma e}\left( r\right) n_{\sigma e}\left(
r^{\prime }\right) }{\left\vert r-r^{\prime }\right\vert }\right) dr\text{\ }%
dr^{\prime }.  \label{eq18}
\end{equation}%
Using the fact that $\Delta (r_{xc}^{\sigma e})\equiv \Delta r_{x}^{\uparrow
\uparrow ,\uparrow \downarrow }+\Delta r_{c}^{e_{i}e_{j}}\ll \left\vert
r-r^{\prime }\right\vert $, this can be expanded to give 
\begin{eqnarray}
E_{xc}\left[ n_{\sigma e}\right] &\simeq &\frac{1}{2}\sum\limits_{\sigma
=\sigma _{1}}^{\sigma _{N_{e}}}\int_{\mathbb{R}^{3}\times \mathbb{R}^{3}}%
\frac{n_{\sigma e}\left( r\right) n_{\sigma e}\left( r^{\prime }\right) }{%
\left\vert r-r^{\prime }\right\vert }\left[ 1-\frac{\ \lambda _{x}\left\vert
\delta r^{\uparrow \uparrow -\uparrow \downarrow }\right\vert +\lambda
_{c}\left\vert r-r^{\prime }\right\vert ^{-\left( N_{e}-1\right) ^{-1}}}{%
\left\vert r-r^{\prime }\right\vert }+O\left( \Delta (r_{xc}^{\sigma
e})^{2}\right) \right] dr\text{\ }dr^{\prime }  \notag \\
&&-\frac{1}{2}\sum\limits_{\sigma =\sigma _{1}}^{\sigma _{N_{e}}}\int_{%
\mathbb{R}^{3}\times \mathbb{R}^{3}}\frac{n_{\sigma e}\left( r\right)
n_{\sigma e}\left( r^{\prime }\right) }{\left\vert r-r^{\prime }\right\vert }%
drdr^{\prime }  \label{eq19}
\end{eqnarray}%
where the first term, being the zeroth-order one, is the uncorrected
classical interelectronic energy, the second indicates the spin and the
Coulomb interaction corrections,  and the third one is associated with
negligible higher order corrections. Omitting $O\left( \Delta
(r_{xc}^{\sigma e})^{2}\right) $ orders, we get 
\begin{equation}
E_{xc}\left[ n_{\sigma e}\right] \simeq -\frac{1}{2}\sum\limits_{\sigma
=\sigma _{1}}^{\sigma _{N_{e}}}\int_{\mathbb{R}^{3}\times \mathbb{R}^{3}}%
\frac{\lambda _{x}\left\vert \delta r^{\uparrow \uparrow -\uparrow
\downarrow }\right\vert n_{\sigma e}\left( r\right) n_{\sigma e}\left(
r^{\prime }\right) }{\left\vert r-r^{\prime }\right\vert ^{2}}dr\text{\ }%
dr^{\prime }-\frac{1}{2}\sum\limits_{\sigma =\sigma _{1}}^{\sigma
_{N_{e}}}\int_{\mathbb{R}^{3}\times \mathbb{R}^{3}}\frac{\lambda
_{c}n_{\sigma e}\left( r\right) n_{\sigma e}\left( r^{\prime }\right) }{%
\left\vert r-r^{\prime }\right\vert ^{\frac{2N_{e}-1}{N_{e}-1}}}dr\text{\ }%
dr^{\prime }.  \label{eq20}
\end{equation}%
For many-electron systems $N_{e}>2$, the exponent $\frac{2N_{e}-1}{N_{e}-1}$
goes rapidly to $2$ and thus the above energy expression becomes 
\begin{equation}
E_{xc}\left[ n_{\sigma e}\right] \simeq -\frac{1}{2}\sum\limits_{\sigma
=\sigma _{1}}^{\sigma _{N_{e}}}\int_{\mathbb{R}^{3}\times \mathbb{R}^{3}}%
\frac{\lambda _{x}\left\vert \delta r^{\uparrow \uparrow -\uparrow
\downarrow }\right\vert n_{\sigma e}\left( r\right) n_{\sigma e}\left(
r^{\prime }\right) }{\left\vert r-r^{\prime }\right\vert ^{2}}dr\text{\ }%
dr^{\prime }-\frac{1}{2}\sum\limits_{\sigma =\sigma _{1}}^{\sigma
_{N_{e}}}\int_{\mathbb{R}^{3}\times \mathbb{R}^{3}}\frac{\lambda
_{c}n_{\sigma e}\left( r\right) n_{\sigma e}\left( r^{\prime }\right) }{%
\left\vert r-r^{\prime }\right\vert ^{2}}dr\text{\ }dr^{\prime }.
\label{eq21}
\end{equation}

\subsection{Dipole potential interpretation}

At this level, a close observation shows that this energy form follows a
multipole expansion behavior without the monopole term being the
zeroth-order term. Inspired by inverse-square behaviors, which have been
largely investigated in quantum physics and related issues \cite{19,20},
these two inverse square potential energy terms could be considered as
potential energies associated with the charge density $n_{\sigma e}\left(
r\right) $ in the magnetic and the electric dipole potentials of the
magnetic and the electric dipole moment densities $m\left( r\right) $ and $%
p\left( r\right) $, respectively. Forgetting about the $-\frac{1}{2}$
coefficient as well as the electromagnetic involved known constants, i.e. $%
\mu _{0}$, $\epsilon _{0}$ which could be absorbed in a redefinition of the
dimension-full factors $\lambda _{x}$ and $\lambda _{c}$, the magnetic and
the electric dipole moments could be expressed as 
\begin{eqnarray}
m\left( r\right) &=&\sum\limits_{\sigma =\sigma _{1}}^{\sigma _{N_{e}}}\int_{%
\mathbb{R}^{3}}\lambda _{x}n_{\sigma e}\left( r^{\prime }\right) \left\vert
\delta r^{\uparrow \uparrow -\uparrow \downarrow }\right\vert dr^{\prime }%
\text{\ }  \label{eq22} \\
\text{\ }p\left( r\right) &=&\sum\limits_{\sigma =\sigma _{1}}^{\sigma
_{N_{e}}}\int_{\mathbb{R}^{3}}\lambda _{c}n_{\sigma e}\left( r^{\prime
}\right) dr^{\prime },  \label{eq23}
\end{eqnarray}%
where $\lambda _{x}n_{\sigma e}\left( r^{\prime }\right) $ corresponds to
the magnetic pole strength density and $\lambda _{c}$ has a dimension of
length. Having these in hand, the associated magnetic and electric dipole
scalar potentials are 
\begin{eqnarray}
\phi _{m}\left( r\right) &\simeq &\frac{m\left( r\right) }{\left\vert
r-r^{\prime }\right\vert ^{2}}\text{\ \ }  \label{eq24} \\
\phi _{p}\left( r\right) &\simeq &\frac{p\left( r\right) }{\left\vert
r-r^{\prime }\right\vert ^{2}}\text{\ }  \label{eq25}
\end{eqnarray}%
from which the expressions for the XC energy functionals read as 
\begin{eqnarray}
E_{x}\left[ n_{\sigma e}\right] &\simeq &\sum\limits_{\sigma =\sigma
_{1}}^{\sigma _{N_{e}}}\int_{\mathbb{R}^{3}}\frac{m\left( r\right) n_{\sigma
e}\left( r\right) }{\left\vert r-r^{\prime }\right\vert ^{2}}%
dr=\sum\limits_{\sigma =\sigma _{1}}^{\sigma _{N_{e}}}\int_{\mathbb{R}%
^{3}}\phi _{m}\left( r\right) n_{\sigma e}\left( r\right) dr  \label{eq26} \\
E_{c}\left[ n_{\sigma e}\right] &\simeq &\sum\limits_{\sigma =\sigma
_{1}}^{\sigma _{N_{e}}}\int_{\mathbb{R}^{3}}\frac{p\left( r\right) n_{\sigma
e}\left( r\right) }{\left\vert r-r^{\prime }\right\vert ^{2}}%
dr=\sum\limits_{\sigma =\sigma _{1}}^{\sigma _{N_{e}}}\int_{\mathbb{R}%
^{3}}\phi _{p}\left( r\right) n_{\sigma e}\left( r\right) dr,  \label{eq27}
\end{eqnarray}%
being, clearly, magnetic and electric potential energies originated from the
magnetic $m\left( r\right) $ and electric $p\left( r\right) $ dipoles
associated with the charge density $n_{\sigma e}\left( r\right) $. It is
denoted that for the case of $\lambda _{x}$, $\lambda _{c}<0$, the
associated dipole potentials (\ref{eq24}) and\ (\ref{eq25}) are attractive
inverse-square potentials and the corresponding XC energy is negative. We
expect this would result in a lowering of the total energy of the real
interacting system (\ref{eq10}). The obtained XC functional result could be
exploited as a possible extension of the LDA functional to unveil new
features for a detailed expression for the XC energy\textbf{\ }via the spin
correction (\ref{eq16}) as it is the case for the correlation correction%
\textbf{\ }(\ref{eq17}). In this way, the analysis of the electric and the
magnetic dipole moments could be useful to evince the main difficulty
associated with explicit expressions. A possible path is to introduce the
inverse square potential scenarios corresponding to the dipole system
behaviors in the Schr\"{o}dinger equation formalism. We believe this will
necessitate the use of high level quantum physics computations. Since such a
potential has been approached from different angles, techniques of extended
physics including Quantum Field Theory could be exploited to explore new
roads to achieve the missing pieces in DFT scenarios \cite{21}.

\section{Conclusion and open questions}

In this work, we have approached the XC energy by reconsidering the study of
the electronic structure calculations of many-electron systems in DFT
scenarios. In particular, the XC energy functional has been dealt with \
using a spatial correction to the electron-electron potential energy to
account for such XC effects. More precisely, by exploiting techniques of the 
$N_{e}$-electron systems in terms of the electron density $n_{\sigma
e}\left( r\right) $ for the obtaining of the ground state energy $E_{e0}$,
we have approached the XC energy $E_{xc}\left[ n_{\sigma e}\right] $ by
means of the interelectronic position corrections emerging from the spin $%
\Delta r_{x}^{\uparrow \uparrow ,\uparrow \downarrow }=\lambda
_{x}\left\vert \delta r^{\uparrow \uparrow }-\delta r^{\uparrow \downarrow
}\right\vert $ and the Coulomb correlation $\Delta r_{c}^{e_{i}e_{j\neq
i}}=\lambda _{c}\left\vert r-r^{\prime }\right\vert ^{-\left( N_{e}-1\right)
^{-1}}$ effects through the Hartree energy $V_{H}\left[ n_{\sigma e}\right] $%
. By considering such corrections, we have obtained approximate expressions
for the exchange $E_{x}\left[ n_{\sigma e}\right] $ and the correlation $%
E_{c}\left[ n_{\sigma e}\right] $ functional energies which could be
interpreted in terms of the electric and the magnetic dipole potential
energies provided by inverse square potential behaviors. The inclusion of
the obtained XC functional energy result in the LDA functional is expected
to allow for more accurate expressions for the XC energy\textbf{. }

This work has offered a particular intuitive spatial approach to the unknown
XC energy, highlighting the need to focus on the interelectronic potential
corrections as a possible route towards new XC energy explicit expressions.
We believe that the implementation of the inverse-square potential, of the
extended inverse-distance potential $v\left( r\right) =\frac{C_{1}}{r}\left[
1+\frac{C_{2}}{C_{1}}\frac{1}{r}+\frac{C_{3}}{C_{1}}\frac{1}{r^{2}}+...%
\right] $, in the generalized Schr\"{o}dinger equation in DFT framework
could open new gates to explore other physical features since such a
potential has been largely investigated in different physical scales
including high energy physics and related topics, e.g. hadronic models and
near black hole horizon behaviors as well as dark matter and dark energy
sectors \cite{21,22,23}.\newline

Data Availability Statement: No Data associated in the manuscript.

\end{document}